\documentclass{article}
\usepackage{bm}
\usepackage[utf8]{inputenc} 
\usepackage[colorinlistoftodos,prependcaption,textsize=small]{todonotes}
\usepackage{graphicx}
\usepackage{wasysym}
\usepackage{url}

\newcommand{\Yes}{{$\CIRCLE$}}
\newcommand{\Par}{{$\LEFTcircle$}}
\newcommand{\No}{{$\Circle$}}

\title{Unified Description for Network Information Hiding Methods}
\author{Steffen WENDZEL$^{1}$, Wojciech MAZURCZYK$^{2,3}$, Sebastian ZANDER$^{4}$}

\begin{document}

This paper is a pre-print of the following publication:\\
~\\

S.\ Wendzel, W.\ Mazurczyk, S.\ Zander: \textbf{Unified Description for Network Information Hiding Methods}, Journal of Universal Computer Science (J.UCS), Vol. 22(11), 2016.\\
~\\

\textit{Please note that the full-text of the final publication can be downloaded under the following URL (open access):}

\url{http://www.jucs.org/jucs_22_11/unified_description_for_network}

\maketitle

\begin{centering}
$^1$ Fraunhofer FKIE, Bonn, Germany\\
$^2$ Warsaw University of Technology, Warsaw, Poland\\
$^3$ FernUniversit{\"a}t in Hagen, Hagen, Germany\\
$^4$ Murdoch University, Perth, Australia\\
\end{centering}

\begin{abstract}
Until now hiding methods in network steganography have been described in arbitrary ways, making them difficult to compare. For instance, some publications describe classical channel characteristics, such as robustness and bandwidth, while others describe the embedding of hidden information. We introduce the first unified description of hiding methods in network steganography. Our description method is based on a comprehensive analysis of the existing publications in the domain. When our description method is applied by the research community, future publications will be easier to categorize, compare and extend. Our method can also serve as a basis to evaluate the novelty of hiding methods proposed in the future.
\end{abstract}


\section{Introduction}

\noindent \emph{Steganography} research determines, describes and evaluates methods that hide information within a medium; \emph{steganalysis} research develops, describes and evaluates methods for the detection and prevention of such methods \cite{petitcolas1999information,Katzenbeisser:2000:IHT:555654}. Steganography has been applied in ancient Greece, in several wars, including World War I and II, and to digital media (digital images, audio files, and digital videos) \cite{petitcolas1999information,Fridrich:IHbook}. \emph{Network steganography} or \emph{network information hiding}, the most recent sub-discipline of steganography, deals with the hiding of information in network traffic \cite{NetStegBook}.

Well over 100 methods for hiding in network transmissions were published since Girling introduced the first methods in 1987 \cite{Girling87}. Wendzel \emph{et al.} clustered these hiding methods in so-called \emph{hiding patterns} and organized these patterns in form of a taxonomy \cite{Wendzel:CSUR}. Hiding patterns are abstract descriptions of hiding methods in a pre-defined format. Using eleven patterns, 109 hiding methods could be described showing how redundant similar ideas were in past research. The introduction of patterns moreover allows to handle hiding methods under a unified term (the pattern) instead of several separate terms introduced by previous research. Mazurczyk \emph{et al.} refined parts of Wendzel's \emph{et al.} work in \cite{NetStegBook}.

On the basis of hiding patterns, a new academic workflow was defined by Wendzel and Palmer for the creativity evaluation of network information hiding methods \cite{StegoCreativity}. The key concept of this workflow is that if a new hiding method cannot be represented by an existing pattern, it comprises higher novelty. The evaluation process of hiding methods in conjunction with hiding patterns is integrated into the traditional peer-review process but requires an author to explain why a hiding method is (or is not) represented by an existing hiding pattern. The approach of \cite{StegoCreativity} fosters the reduction of terminology inconsistencies as new publications will be aligned to existing pattern terminology and the improved peer review process eases spotting any inconsistencies.

The aforementioned publications emphasize on the categorization of hiding methods, the reduction of inconsistent terminology and redundant ideas, and the evaluation of the novelty of hiding methods.

\begin{figure*}[!th]
\centering
  \includegraphics[width=0.99\textwidth]{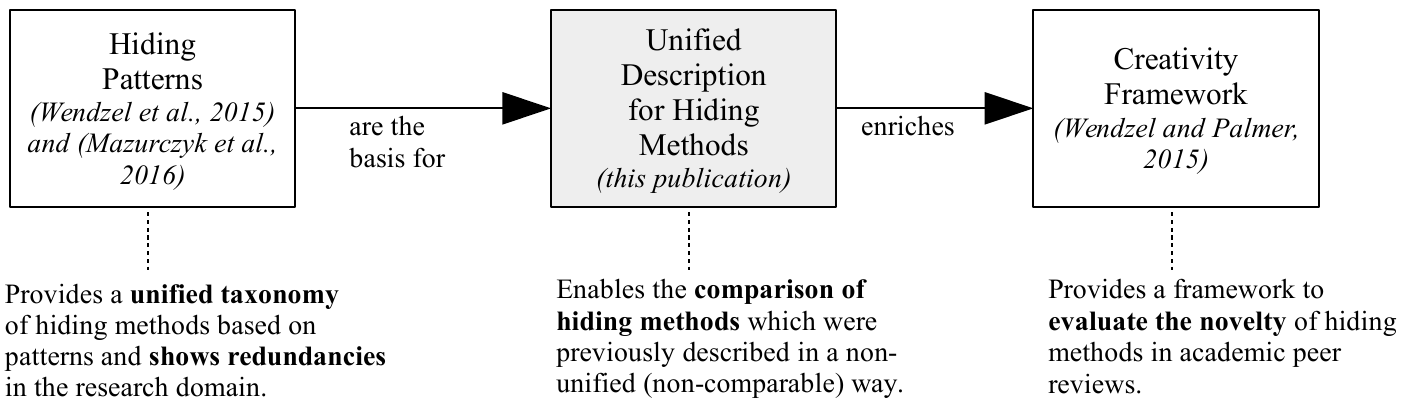}
  \caption{Contribution of this work: While hiding patterns serve as a basis for this new publication, it provides its own contribution by enabling the structured comparison of research work on hiding methods and it can be also used in conjunction with the creativity framework.}
  \label{Fig:contribution}
\end{figure*}

After analyzing 131 hiding method descriptions from 74 publications published between 1987 and 2015, we found relevant inconsistencies in the \emph{description} of hiding methods. In particular, we noticed large differences in the evaluations and technical descriptions between the publications. While for some hiding methods, the channel capacity is described, other publications focus solely on the embedding process, the application scenario or other aspects. Moreover, the descriptions of hiding methods even vary \emph{within} some of the papers. We also found that some papers \emph{combine} the evaluation and description for several hiding methods. For example, some publications discuss the overall throughput of multiple hiding methods instead of discussing the different channels separately. These non-unified descriptions make it difficult to compare publications that propose new hiding methods and to evaluate each described hiding method separately.

\textbf{Contribution.}
In this paper, we introduce a method for a unified description of hiding methods in network steganography. Figure~\ref{Fig:contribution} visualizes our contribution in the context of previous work. The existing hiding patterns describe in a generic way how hidden information is signaled, and the creativity framework provides a way to reduce redundant research outcomes and to evaluate whether a proposed hiding method is actually new, or not.

This paper fills a missing gap in the previous work by introducing a multi-facetted and detail-rich description for hiding methods. When applied by the scientific community, our unified description method will enable the easy comparison, categorization, and evaluation of the hiding patterns. Our description method can also serve as a basis for the creativity framework and will enrich the evaluation process of the same.

\textbf{Outline.}
Section~\ref{Sect:Fundamentals} introduces fundamentals and related work. Section~\ref{Sect:UnifDescMethodOverview} provides an overview of the unified description method for hiding methods. The description method is split into three main categories, which are covered in sections~\ref{Sect:GeneralInf}--\ref{Sect:Countermeasures}. We provide two exemplary descriptions in section~\ref{Sect:Examples}. We discuss results of our literature analysis in section~\ref{Sect:LitSurvey}. Section~\ref{Sect:Creativity} explains how the unified description can be used in combination with a creativity framework to evaluate the novelty of newly proposed hiding methods. We provide a conclusion in section~\ref{Sect:Discuss}.

\section{Fundamentals}\label{Sect:Fundamentals}

Patterns are used in various sciences and can be seen as abstract descriptions of recurring designs. By definition, \emph{patterns} represent a design (or solution) to a re-occurring problem in a given context. In network steganography, hiding patterns describe how to use use a method (solution) to hide data (problem) in network traffic (context) \cite{Wendzel:CSUR,StegoCreativity}.
In other words, a pattern describes a hiding method in an abstract way. For instance, a pattern can describe that bits can be hidden in the least significant bits (LSB) of network header fields but it does not cover details on how this will be achieved within a particular network protocol. 
Patterns can also contain a description of their relation to other patterns, forming not only a classification for hiding methods but a taxonomy of hiding methods \cite{Wendzel:CSUR}.

Classifications and taxonomies have a long history and were already applied in other scientific domains. In ancient Greece, Aristotle wrote his \emph{Categories} which contain a taxonomy of living things. In 1735, Linnaeus published his \emph{Systema Naturae}, providing a taxonomy of the Animal, Plant and Mineral Kingdoms. 
In the 19th century, Danish curator Thomsen introduced the \emph{three-age system} to categorize archeological artifacts into three main ages (Stone, Bronze and Iron age) based on industrial stages \cite{Three-age-system}. Manktelow provides an overview with additional examples of taxonomy development throughout the history of Biology in \cite{Manktelow:HistoryTaxonomy}.

While taxonomy and categorization can be applied at large scale, such as for the categorization of the Animal Kingdom, it can be also applied to smaller areas, such as network information hiding methods. Our paper is not a replacement for \cite{Wendzel:CSUR} which introduced a taxonomy for hiding methods. Instead, it significantly extends the general classification of hiding patterns in \cite{Wendzel:CSUR,NetStegBook} and supports the workflow in \cite{StegoCreativity} by introducing a description for hiding methods on a detailed level.

\section{Unified Description Method}\label{Sect:UnifDescMethodOverview}

We analyzed the literature describing 131 hiding methods published since 1987. The analysis revealed that the descriptions of hiding methods in the related publications differ significantly regarding their provided information. To improve this situation, we designed a unified description method.

\subsection{Applicability to Scientific Work}

The unified description method can be directly applied to structure new scientific papers. This way, authors which present hiding methods make it easy for other researchers and reviewers to compare the new hiding method with existing hiding methods.
Our unified description can be also combined with the `creativity framework' \cite{StegoCreativity} to evaluate the \emph{novelty} of a proposed hiding method by applying concepts of creativity research during the academic peer review process. By combining the creativity metric with our unified description method, both, the presentation of a hiding method and the underlying research novelty can be compared in a process that is unified, reasonable and re-constructable.

\subsection{Overview of the Description Method}

\begin{figure}
  \centering
  \includegraphics[width=0.6\textwidth]{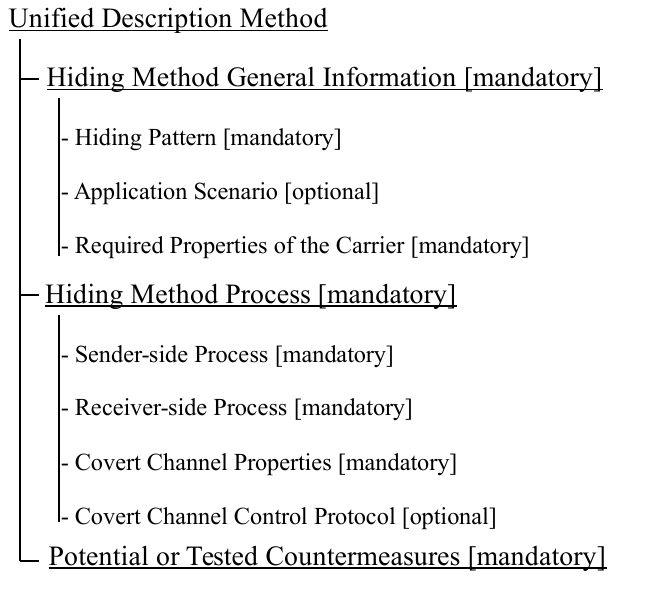}
  \caption{Overview of the description method's structure.}
  \label{Fig:overview}
\end{figure}

Our description of hiding methods is split into three categories, namely \emph{general information about the hiding method}, \emph{description of the hiding process}, and \emph{potential or tested countermeasures}. The first two categories comprise sub-categories and each (sub-)category can be a mandatory or optional. Figure~\ref{Fig:overview} provides an overview.

The category `hiding method general information' consists of a link to an existing hiding pattern and a detailed description of the hiding method. It also includes a discussion of the application scenario and requirements of the carrier. The category `hiding method process' is split into four parts: the sender-side and the receiver-side description of the hiding method, the details of the covert communication channel, and the description of an associated covert channel control protocol (if applicable). The third category discusses both potential and evaluated countermeasure, including those that detect, limit or prevent the particular hiding method's use. The following three sections will explain all categories.

\section{Hiding Method General Information}\label{Sect:GeneralInf}

This section describes the general attributes of the steganographic method. These attributes include the hiding pattern that the method belongs to, the considered and potential application scenarios, and general requirements for the carrier.

\subsection{Hiding Pattern [mandatory]}

In \cite{StegoCreativity}, we proposed that when a new hiding method is to be published, it should be assigned to a particular pattern or provide evidence why it does not match any of the existing patterns. In case the proposed hiding method does not match any existing pattern, a new pattern can be created. As the existing patterns are based on many hiding methods invented since 1987, it is likely that a new hiding method can be represented by an existing pattern. In the less likely case that no existing pattern fits, the authors of the new method must provide a detailed explanation of a new pattern that underpins the novelty of the hiding method they propose. The consequence of a new pattern is an extension of the existing pattern catalog. %

If the authors decide a hiding method can be represented by an existing pattern, the pattern should be stated including the whole hierarchical path of the pattern in the pattern hierarchy (including sub-patterns). The hiding pattern hierarchy is described in detail in \cite{Wendzel:CSUR} and an on-line version is available under \emph{https://ih-patterns.blogspot.com}.

A pattern name including the path within the hierarchy is the complete path from the root node of the hierarchy to the leaf that represents the pattern. For instance, a hiding method that modifies the least significant bit (LSB) of a header element would be represented by the ``LSB'' pattern and the full path of the hierarchy would be:

\begin{verbatim}
Network Covert Storage Channels
`-- Modification of Non-Payload
    `-- Structure Preserving
        `-- Modification of an Attribute
            `-- Value Modulation
                `-- Least Significant Bit (LSB)
\end{verbatim}\normalsize

For each element of the hierarchy, it should be explained briefly why the described hiding method belongs to the element, e.g.\ why the new method is a covert storage channel (and not a timing channel), why it preserves instead of modifies the structure of a PDU, why it is an attribute modification, value modulation, and LSB-based method. Using the hierarchy-based explanation, every reader who has knowledge of the pattern hierarchy can easily follow and verify the argumentation of an author.

\subsection{Application Scenario(s) [optional]}

This category describes the application scenario for which a hiding method was developed. It helps to identify novel application scenarios and makes it easier to compare different methods as some hiding methods may have application-specific limitations. Such methods may not be usable in other scenarios. For example, hiding methods developed for breaking anonymization \cite{zander08Clockskew} may provide only small throughput making them unusable for general-purpose communication.

Many hiding methods were developed for general-purpose communication, i.e.\ the passing of secret messages between two or more parties. Typically this application is motivated by the existence of an adversarial relationship between different groups, such as government agencies versus criminal or terrorist organizations or dissenting citizens versus their governments. Other existing hiding methods are tailored to the case of hackers or corporate spies whose aim is to covertly control compromised systems or ex-filtrate data from compromised systems. Similarly, malware, such as computer viruses or worms, can use hiding methods to spread undetected, to ex-filtrate data, or for covertly exchanging information (e.g.\ execute brute-force attacks on cryptosystems \cite{white89}). Indeed, this rising trend has been recently confirmed by many real-life examples of information hiding-capable malware \cite{Mazurczyk2015}.

On the other hand, there are hiding methods that were developed for very specific contexts. Some hiding methods were developed for breaking anonymization, for example Murdoch \emph{et al.} developed methods to reveal servers hidden inside anonymization networks \cite{murdoch05b, zander08Clockskew}. Other hiding methods were developed for transmitting authentication data, for example to allow authorized users to access open firewall ports while presenting these ports as closed to all other users (``port knocking'') \cite{degraaf05}. Another type of hiding methods were designed for packet/flow traceback or watermarking -- techniques used for linking different observable instances of network packets or flows in scenarios where packet contents cannot be used for linking \cite{houmansadr09}. Another specific application are hiding methods developed for cheating in on-line games \cite{murdoch04}. 

In case a hiding method is for general-purpose communication, no comprehensive description is needed. However, for methods that were developed for a specific application in a specific context, the application scenario should be described in detail. Also, new application scenarios should be described in more detail than well-known scenarios.

\subsection{Required Properties of the Carrier [mandatory]}

This category is used to specify the properties of the carrier that the hiding method requires. It should describe whether the hiding method is limited to a certain protocol (or a service) as carrier or whether it can be used with several or even many different carrier protocols/services.

If the hiding method is tied to a single carrier protocol, the description must specify the protocol and describe the specific protocol features that are used by the hiding method. If a hiding method works with a set of carrier protocols, the description must specify the protocols and the protocol features the hiding method relies on. If a hiding method depends on certain protocol features that are common to a large number of protocols, the description must list the features and describe them. For truly generic hiding methods that work with all kinds of carrier traffic the description may be short; however, in our experience such generic hiding methods are rare. 
 
For hiding methods that are not only tied to certain protocols or protocol features, but also require certain operational conditions, these conditions must also be described. For example, a method that hides information by intentionally introducing packet losses assumes that packets of the carrier can be discarded and also it can only blend in with the normal traffic if there is natural packet loss \cite{kraetzer06}; hence, the possibility and occurrence of natural packet loss is an operational condition for this hiding method.

\section{Hiding Method Process}\label{Sect:HidProc}

This section covers the categories which describe the actual process of the hiding method, including the embedding and the extraction of hidden data, as well as the channel properties and a potentially present control protocol.

\subsection{Sender-side Process [mandatory]}\label{SSP}

This category describes the embedding process performed by the covert sender to hide secret data. It must be explained whether the sender is a centralized host/process or distributed. In the classical scenario, one sender transfers secret data to one receiver (the sender-to-receiver relationship is `1:1'). However, other scenarios are also possible and they depend on the specific context in which a covert transmission is performed or it can be a characteristic feature of the carrier utilized for hidden data exchange. In case of covert channel overlay networks, it is imaginable that one sender broadcasts the covert data to multiple receivers (`1:m'). In case of a distributed sending system, there may be $n$ hosts forming one logical sender that transfers data to one or multiple receivers (`n:1' and `n:m' relations). For instance, if the source address of a receiver indicates a hidden bit, two senders can be used to transfer a message of zero and one bits to a receiver.

The location of the covert data can be also centralized or distributed depending on whether the hidden data is `inserted' into a single carrier (or a subcarrier) or it is distributed across several carriers (subcarriers). This means that the covert data can be embedded, e.g.\ into one particular part of a packet or into multiple areas of a packet, but it can be also distributed among different flows \cite{NetStegBook}.

It must be also specified whether the steganographic method generates its own cover traffic or whether data is hidden in third-party cover traffic. In case the hiding method is responsible for the cover traffic generation then a description of this process must be included here.

\subsection{Receiver-side Process [mandatory]}
This category describes the recognition and extraction process of the covert data at the receiver-side. Similar to the sender-side process description, the secret receiver can be also centralized or distributed and the same considerations apply here (see section \ref{Sect:HidProc}.1). 
If the receiver is a distributed system, it should be explained how the covert data is extracted from the hidden data carrier and how it is finally merged. 

\subsection{Covert Channel Properties [mandatory]}
In this category the considered hidden communication scenario(s) for the particular steganographic method should be described and it should be indicated whether the created covert channel is direct or indirect. Moreover, four characteristic features of the information hiding technique should be analyzed. This will allow to describe properties of the created covert channel. All above mentioned attributes are explained below.

\begin{figure*}[!th]
  \centering
  \includegraphics[width=0.99\textwidth]{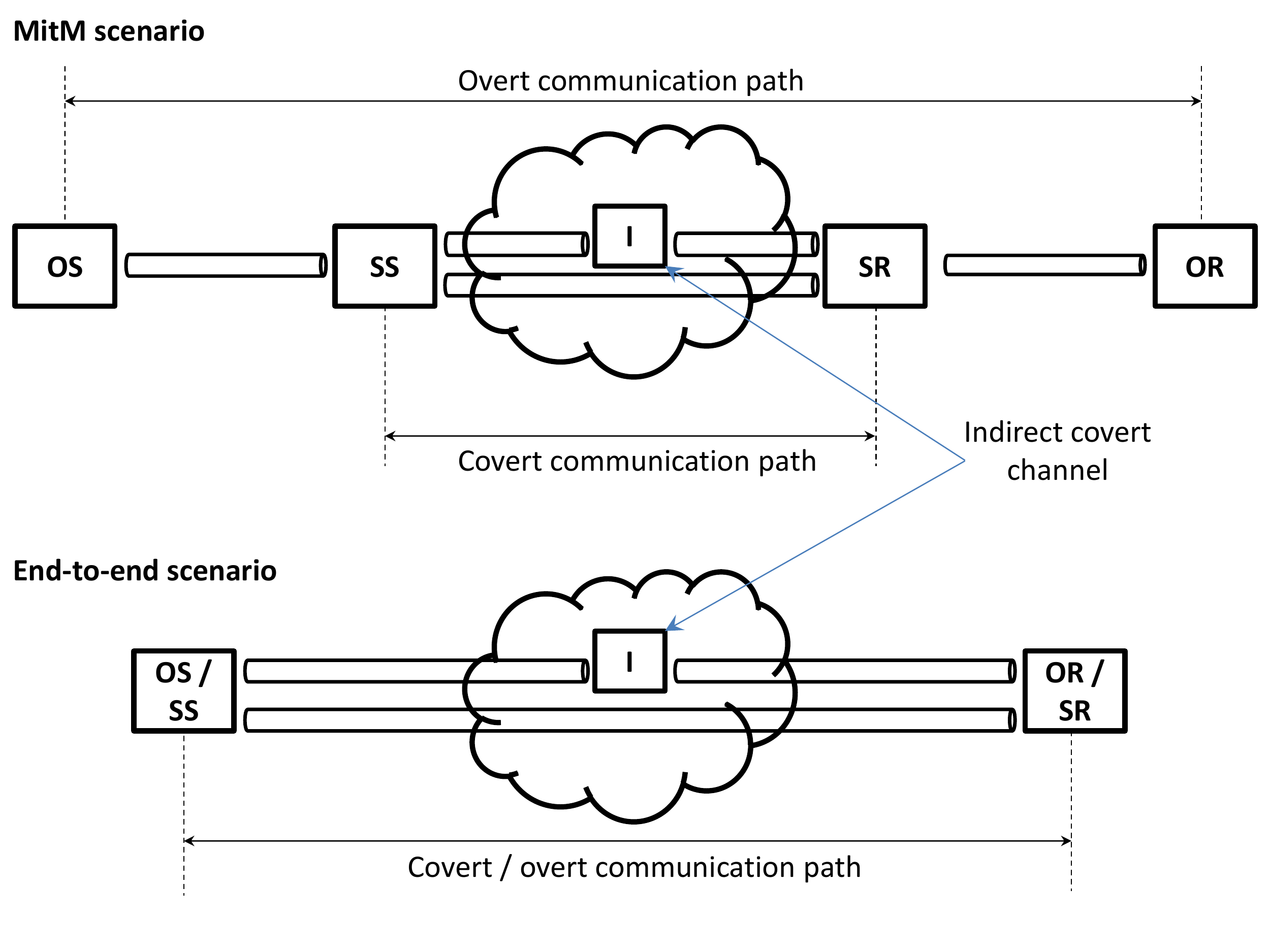}
  \caption{Hidden communication scenarios (OS -- overt sender, OR -- overt receiver, SS -- secret sender, SR -- secret receiver, I - intermediate node)}
  \label{Fig:scenarios}
\end{figure*}

For network steganography, two main possible communication scenarios may be considered, as illustrated in Fig. \ref{Fig:scenarios}. The first scenario, i.e.\ end-to-end scenario, is the most common: the secret sender and the secret receiver perform overt communication while simultaneously exchanging covert data. In this case the overt communication path is equal to the covert path. In the second scenario, i.e.\ Man-in-the-Middle (MitM) scenario, only a part of the end-to-end overt communication path is used for the hidden communication, as a result of actions undertaken by intermediate covert nodes. Therefore, the overt sender and overt receiver are, in principle, unaware of the steganographic data exchange. Obviously hybrid scenarios are also possible where the overt sender/receiver serves as secret sender/receiver but the other covert party is located in some intermediate node.

Next, it should be indicated whether the covert channel is \textit{direct} or \textit{indirect}, i.e.\ whether the overt traffic flows directly from the secret sender to the secret receiver or via one or multiple intermediaries. In case of a direct channel the overt traffic that contains the covert data flows directly from the secret sender to the secret receiver (who both can act as middlemen). A covert channel is indirect when the secret sender does not send covert data directly to the secret receiver (or a destination downstream of the secret receiver). Instead, the secret sender transmits the covert data to an intermediate host which then unknowingly forwards (due to the functions of the overt traffic protocol) the covert data to the secret receiver. This means that there are two flows of the overt traffic conveying the covert data, i.e.\ the first flow is between the secret sender and an unwitting intermediary and the second flow is between the intermediary and the secret receiver. Indirect covert channels provide an increased stealthiness as a warden does not observe a direct flow of information between the covert sender and receiver. On the other hand they are typically harder to implement and have a smaller capacity than direct channels. In case of an indirect channel the requirements on the intermediate need to be described.

For example, Rowland \cite{Rowland} proposed an indirect channel that exploits the TCP three-way-handshake. Instead of sending a TCP SYN segment with an ISN containing covert data directly, the secret sender sends the TCP SYN segment to the intermediary (a bounce host) with a spoofed IP source address set to the intended destination. The intermediary then sends a SYN/ACK or SYN/RST to the secret receiver with the acknowledged sequence number equal to the ISN+1. The secret receiver decodes the hidden information from the ACK number (ACK$-$1).

The classic characteristics of any information hiding method as mentioned by Fridrich in \cite{Fridrich:IHbook} are: \textit{steganographic bandwidth} (the amount of secret data that can be sent per unit time when using a particular method), \textit{undetectability} (defined as the inability to detect a steganogram within a certain carrier), and \textit{robustness} (defined as the amount of alteration covert data can withstand without the secret data being destroyed). As proposed in \cite{StegoCost} another attribute is the \textit{steganographic cost}, which describes the degradation or distortion of the carrier caused by the application of the steganographic method. All these attributes should be described, if feasible. 
Especially for novel methods and for the description of third party tools, a comprehensive description of all four characteristics is hardly feasible. The author of a paper may have no access to implementations of these countermeasures, to testbeds in which countermeasures could be evaluated, or may have no knowledge of all existing countermeasures. %
Since the detectability issues are also discussed in the category `Potential and Tested Countermeasures' only a brief description or reference to that category is required here. 

\subsection{Covert Channel Control Protocol [optional]}\label{Sect:ControlProtocols}

Several hiding methods utilize so-called covert channel control protocols. Covert channel control protocols embed small protocol headers providing several features including reliable data transfer (by introducing sequence numbers and ACKs), peer discovery and dynamic overlay routing (between steganographic peers), session management for covert transactions, adaptiveness and several features of application layer protocols (e.g.\ file transfer features) \cite{MPSurvey}. 
If utilized by a hiding method, both the design, implementation and features of a covert channel control protocol must be described in this category. Otherwise, the category can be left empty.

\section{Potential or Tested Countermeasures}\label{Sect:Countermeasures}

This category comprises no sub-categories. It should describe potential as well as tested countermeasures available against the hiding technique. There are three types of countermeasures that can be applied against the covert channel created by a hiding method: elimination, limitation, and detection \cite{zander07}. Not all of these three types may be applicable for a particular hiding method, for example some covert channels cannot be eliminated. The description must contain a discussion which countermeasures are applicable and which are not applicable including a justification. Applicable countermeasures should then be described in detail.

Elimination means the covert channel created by a hiding method can be eliminated completely. For example, a method that hides data in unused header fields or padding bytes can be eliminated completely by a traffic normalizer that sets the unused header fields or padding bytes to a default value (e.g.\ zero). Some hiding methods cannot be eliminated, for example covert channels in on-line game protocols \cite{zander08MPGames}. If a covert channel can be eliminated, the description should include a discussion on how the elimination works and any possible limitations. The description should also include side-effects of the elimination process. For example, if a covert channel in a header extension can be eliminated by removing the header extension from the packets, then the description should include a discussion of the impact on the protocol functionality.

Limitation means the covert channels can be perturbed, for example by introducing noise, so that its capacity is greatly reduced and the channel effectively becomes useless. Limitation usually has side-effects on the carrier, so there is a trade-off between reducing a covert channel's capacity to a small value and not significantly impacting on the carrier protocol. The description should include a discussion on whether a channel's capacity can be limited and the impact on the carrier should be characterized. If a channel can be eliminated then a description of a limitation method is optional.

Elimination or capacity-limitation are active countermeasures that require a warden to manipulate the network traffic (active warden) \cite{Fisk2003}. However, having an active warden may not be possible in every scenario. 

Detection means the existence of the covert channel inside the carrier can be detected, which requires a passive warden to observe the traffic. Usually detection mechanisms are based on some characteristics that can be observed, and the characteristics for traffic with covert channels are different from the characteristics of regular traffic (traffic without covert channels). While detection itself is passive, it can be coupled with active measures such as targeted blocking, elimination or limitation where the warden can manipulate suspicious traffic with more impunity. The description should include whether the channel can be detected and outline the detection method. If the channel is impossible to detect, the description should provide a justification why it is undetectable. If a detection method is introduced, the proposed characteristics for identifying the covert channel must be defined.

If an evaluation of the proposed elimination, limitation or detection method(s) has been conducted, the description should summarize the evaluation scenario(s) and results. Ideally, an evaluation is done under realistic conditions, e.g., in real networks using realistic traffic, but in practice this is not always feasible. The description of the evaluation should point out any such limitations. 

Another type of countermeasure is to change the specification of a network protocol to prevent its use as carrier in the future. For example, a network protocol prone to covert channels could be revised and updated with a newer version less prone to covert channels. In many cases this may not be realistic, as widely deployed protocols cannot be changed easily. However, in cases where an updated protocol could be realistically deployed, this should be described.

The description should also discuss whether the warden can be a single entity (centralized warden) or has to be multiple distributed instances (distributed warden), and whether the warden has to keep flow state (stateful warden) or can operate without flow state (stateless warden).

\section{Exemplary Descriptions}\label{Sect:Examples}

We now present two exemplary descriptions. The first description is for a covert timing channel, while the second description is for a covert storage channel.

\subsection{Example 1: Inter-packet Timing Method}\label{Example1}

In this example we describe a specific steganographic method for hiding information in inter-packet timings. This method or channel is also referred to as model-based inter-packet gap channel and was proposed by Gianvecchio \emph{et al.} \cite{gianvecchio08}.

\subsubsection{Hiding Pattern}

As the covert signaling utilizes the timing of network packets, the method belongs to the `Network Covert Timing Channel' pattern. In particular, the method falls under `Inter-arrival Time' \cite{Wendzel:CSUR}. The full path in the pattern hierarchy is:

\begin{verbatim}
Network Covert Timing Channels
`-- Inter-arrival Time Pattern
\end{verbatim}
\normalsize

\subsubsection{Application Scenario}

The method can be used for general-purpose covert communication between a covert sender and a covert receiver or between a group of covert parties depending on whether the carrier is unicast or multicast.

\subsubsection{Properties of the Carrier}

The method only requires that the carrier consists of packetized data, such as network-layer packets, whose timing can be manipulated. The method assumes that there is sufficient noise in the timing of packets by senders and along the path, so that manipulations of timings are not immediately suspicious. 
While the method was proposed and evaluated for HTTP in \cite{gianvecchio08}, it is not limited only to this protocol. However, some carrier protocols are more suitable than others. Since the encoding destroys any dependence between the inter-packet times of successive packets, it is 
best used with carriers that already have independent inter-packet times \cite{zander11stealthier}.

\subsubsection{Sender-side Process}

The embedding process involves fitting a model to the inter-packet time distribution of regular traffic and then using the model to generate covert channels with identical distribution (for details see \cite{gianvecchio08}). There is usually a single sender process
that embeds the covert channel in a single carrier. Note that a single carrier can be multiple 
traffic flows.

In the original work the covert sender generated the overt traffic \cite{gianvecchio08}. However, the method can be also applied to embed the covert channel into existing network traffic at the cost of increasing the latency of the overt traffic \cite{zander11stealthier}.

\subsubsection{Receiver-side Process}

The extracting process running on the single receiver decodes the covert bits from the observed 
inter-packet times of the single carrier as described in \cite{gianvecchio08}.

\subsubsection{Covert Channel Properties}

The method can be used in the end-to-end scenario, where covert sender and receiver are also the overt sender and receiver, and in a MitM scenario, where covert sender and receiver are placed between the actual sender and receiver (as well as in hybrid scenarios). The method creates a direct channel between covert sender and receiver(s).

The steganographic bandwidth depends on the channel noise and the packet rate of the carrier traffic. Gianvecchio \emph{et al.} measured capacities of 5--20 bits per second in their experiments. Note that in practice the goodput is likely smaller as part of the capacity will be used by a control protocol to provide reliable transport.

The channel is hard to detect only if the regular traffic has uncorrelated inter-packet times which is largely the case when HTTP is used as a carrier (as in \cite{gianvecchio08}). Otherwise the channel can be detected with metrics that can measure the dependency of inter-packet times \cite{zander11stealthier}.

The channel is robust against typical network packet timing noise. If an active warden can manipulate packet timings without impunity, the capacity of the channel would be severely reduced up to a degree where the channel would be practically eliminated. %

Measurements regarding the steganographic cost were not provided by the authors as the concept of steganographic cost had not been introduced at that time. The steganographic cost depends on the abovementioned channel characteristics and the carrier traffic. In general, the more severely delays are perturbed in overt traffic, the higher the steganographic cost.

\subsubsection{Control Protocol}

Gianvecchio \emph{et al.} \cite{gianvecchio08} only describe the ``physical layer'' of the covert channel (encoding/decoding of bits) and do not mention a control protocol. 

\subsubsection{Countermeasures}

The covert channel can be limited or even practically eliminated by introducing timing noise, either at the sender or in the network \cite{Fisk2003}. Depending on the carrier this may introduce unwanted side-effects though, for example, it may add additional latency to the carrier application's traffic.

The covert channel mimics the distribution of inter-packet times of the normal traffic. This makes the channel hard to detect if the normal traffic has uncorrelated inter-packet times \cite{gianvecchio08}. However, for applications that have correlated inter-packet times, the channel can be detected with metrics that can measure the dependency of inter-packet times \cite{zander11stealthier}.

\subsection{Example 2: DHCP Number of Options Storage Method}\label{Example2}

We now discuss the description of a covert storage channel. Rios \emph{et al.} presented several DHCP-based covert channels, of which one hides information by changing the number of DHCP options in a DHCP packet \cite{RiosEtAlDHCPHIDE}.

\subsubsection{Hiding Pattern}

As the modification of DHCP options represents the modification of a storage attribute, the method falls under `Network Covert Storage Channels'. The DHCP options are part of the DHCP header and thus, a `Modification of Non-Payload'. They are also `Structure Modifying' as the header structure is extended when DHCP options are embedded. The signaling of the hidden information is performed in a way that a sequence of objects (DHCP options) is utilized (`Sequence Pattern') and, in particular, the number of options represents the hidden information itself (`Number of Elements Pattern') \cite{Wendzel:CSUR}. The full path in the pattern hierarchy is:

\begin{verbatim}
Network Covert Storage Channels
`-- Modification of Non-Payload
    `-- Structure Modifying
        `-- Sequence Pattern
            `-- Number of Elements Pattern
\end{verbatim}
\normalsize

\subsubsection{Application Scenario}

Rios \emph{et al.} discuss a potential application in a data exfiltration scenario \cite{RiosEtAlDHCPHIDE}: Alice, having privileged access to an embassy network, needs to receive information from Bob, but the direct communication between Alice and Bob is forbidden and the Internet-based communication between them would be suspicious. Bob visits the embassy and transfers network messages to Alice. He embeds hidden information within the non-blocked DHCP protocol using the local network. The application scenario foresees only an uni-directional communication, i.e.\ no backwards channel from Alice to Bob, but in general the channel could be bi-directional.

\subsubsection{Properties of the Carrier}

The DHCP protocol must be allowed, i.e.\ not administratively prohibited by a network security policy (e.g.\ blocked by a switch or layer-2 firewall). As the intended transfer of hidden information is uni-directional, i.e., only from Bob to Alice, Alice is not required to be able to send over the carrier herself. 

The hiding method is protocol-specific and can only be applied to the DHCP protocol. %
To apply the technique, the network must not block particular DHCP options and the encoding of hidden information must be performed in a way that for all encodable symbols, the resulting DHCP packet is still transferable over the carrier.

\subsubsection{Sender-side Process}

The secret sender generates its own overt traffic. The sender-side process embeds a hidden symbol by adding the number of DHCP options the symbol requires for its encoding to the DHCP packet. At least two options must be embedded due to the DHCP standard, for example, if the symbols are `A'--`Z', the symbol `A' requires two options already \cite{RiosEtAlDHCPHIDE}. The symbol `Z' would require 27 options, which is likely to raise suspicion \cite{RiosEtAlDHCPHIDE} and may be blocked by firewalls. Each symbol to be transmitted must be encoded in a separate packet.

Reliability is not implemented directly -- instead the \emph{recovery mechanisms provided by DHCP against packet loss} are exploited \cite{RiosEtAlDHCPHIDE}.

\subsubsection{Receiver-side Process}

The receiver observes DHCP messages sent by the covert sender and counts the number of embedded DHCP options. The number of DHCP options represents the hidden symbol. The decoding is performed separately for each DHCP packet and the received symbols are combined to reassemble the transmitted message.

\subsubsection{Covert Channel Properties}

The method works in an end-to-end communication scenario. It cannot be used in a MitM scenario due to the properties of DHCP (broadcast messages that are limited to one subnet). 
The channel is a direct channel.

The bandwidth of the channel depends on the number of DHCP packets sent per second. In general, the channel can transfer as many symbols as packets per second.

The detectability of the channel increases with the number of symbols encoded per second and with the size of the encoded symbol. The detectability of a message transfer could be improved by encoding the most frequently used symbols with shorter messages (i.e.\ apply a Huffman coding).

Robustness of the covert communication is provided by the use of DHCP's recovery mechanisms.

Measurements regarding the steganographic cost were not provided by the authors as the concept of steganographic cost had not been introduced at that time. In general, the distortion of the used carrier (e.g.\ the Ethernet environment) is minimal as long as the number of DHCP packets does not influence the network's performance in a recognizable manner.

\subsubsection{Control Protocol}

No control protocol was described in \cite{RiosEtAlDHCPHIDE}. Several existing control protocols for covert storage channels are surveyed in \cite{MPSurvey} but are generally designed for a bi-directional communication. As stated in \cite{RiosEtAlDHCPHIDE}, a bi-directional communication over the covert channel is feasible, so bi-directional control protocols could be integrated.

\subsubsection{Countermeasures}

The easiest way to eliminate the method is to prevent the use of the DHCP protocol or DHCP packets with more than two options. However, this solution may not be practically applicable as the DHCP protocol is essential in most networking environments.
A traffic normalizer that deletes uncommon or redundant DHCP options would be a better solution but it may eliminate actually required protocol functionality.

Another potential countermeasure would be to limit the number of DHCP packets per second. This approach reduces the channel's performance as each symbol must be encoded in a separate packet.
As the number of DHCP packets per second is unlikely to be high during regular transmissions, a statistical analysis will probably allow an accurate detection of the steganographic method.

Rios \emph{et al.} state that large DHCP packets, i.e.\ those with many options, may raise suspicion~\cite{RiosEtAlDHCPHIDE}. DHCP packets with an unusual large number of embedded options can likely be detected with simple intrusion detection rules.

\section{Literature Analysis}\label{Sect:LitSurvey}

\begin{figure}[!th]
\centering
  \includegraphics[width=0.99\textwidth]{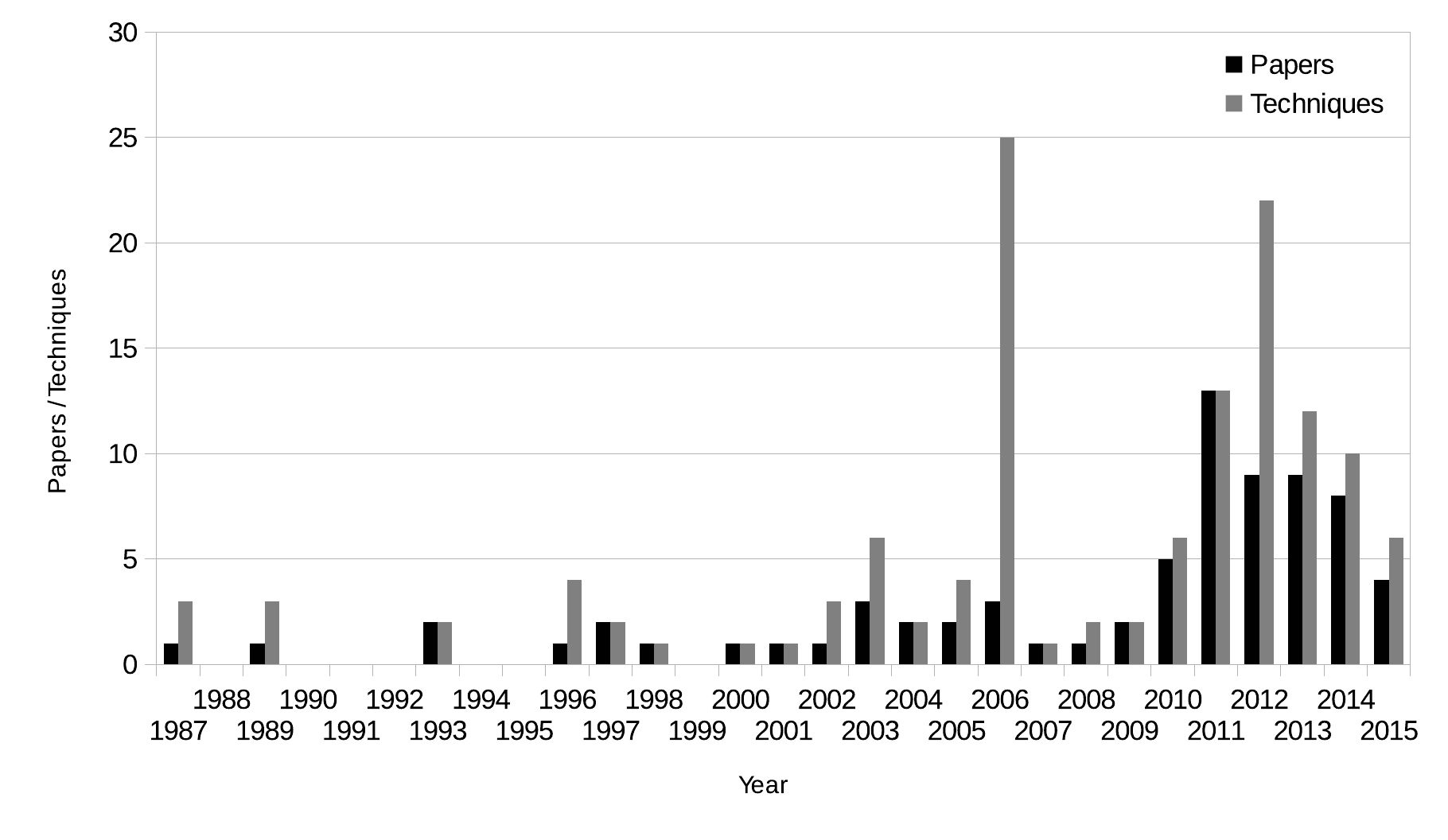}
  \caption{Analyzed publications that present hiding methods (per year).}
  \label{Fig:PubsPerYear}
\end{figure}

We analyzed 74 publications published between 1987 and 2015. 
Figure~\ref{Fig:PubsPerYear} shows the analyzed publications per year. In early years, only few papers on network covert channels were published. The number of papers grew over time due to the increasing popularity of the topic. As some papers describe more than one hiding technique, the number of analyzed hiding techniques (n=131) is sometimes larger than the number of publications, which is also shown in Fig.~\ref{Fig:PubsPerYear}.

\subsection*{\textbf{Finding 1: Several papers lack fundamental attributes}}

\begin{figure}[!th]
\centering
  \includegraphics[width=0.99\textwidth]{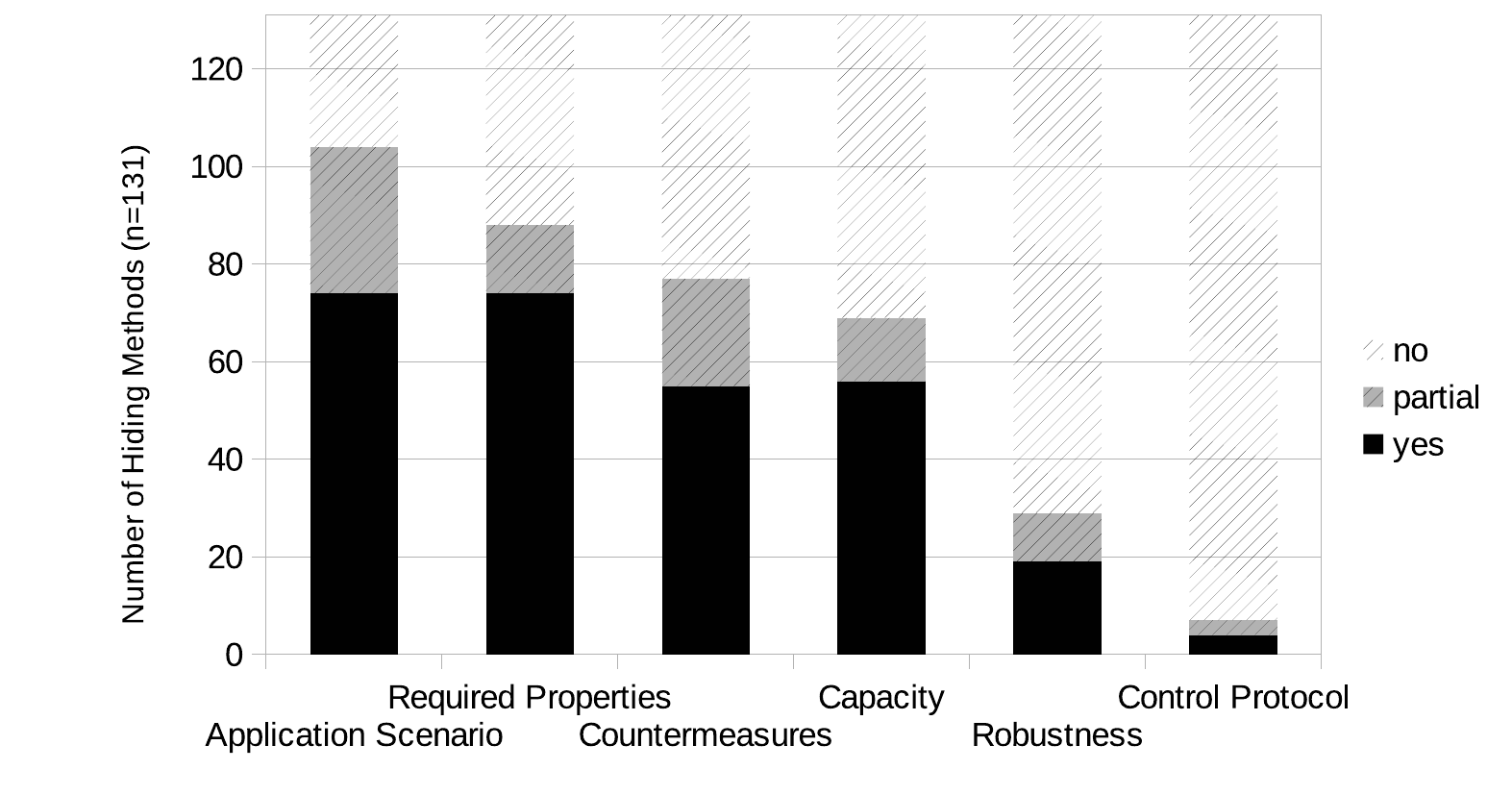}
  \caption{Presence (fully or partially) of selected attributes in the publications.}
  \label{Fig:coverage}
\end{figure}

As already stated in the introduction, our analysis shows that publications on hiding methods present varying subsets of attributes. Figure~\ref{Fig:coverage} provides an overview of the present attributes for all described hiding methods of the analyzed papers. When an attribute's description is classified as `partial', the authors provided some aspects but lack other fundamental aspects of the particular attribute. The comprehensive description of an attribute was marked with `yes' (fully present).

Out of the 131 described hiding methods, the application scenario was provided for 78\% of them (74 fully, 30 partially). The required properties for the hiding method were fully described for 74 and partially for 14 hiding methods (combined 67\%). For some of the techniques, the authors provided countermeasures: 55 contained a full description with evaluation of at least one countermeasure and for another 13 techniques, possible countermeasures were at least briefly discussed (combined 58\%).

The channel capacity was evaluated for 52\% of the hiding methods (56 fully, 13 partially; both values also including throughput and bitrate measurements). The robustness of the proposed hiding methods was discussed for only 22\% of the hiding methods (19 fully, 10 partially). Control protocols are not part of most hiding methods and for this reason only described for 5\% of the hiding methods (4 fully, 3 partially).

\subsection*{\textbf{Finding 2: Attribute coverage changed over time}}

\begin{figure}[!th]
\centering
  \includegraphics[width=0.99\textwidth]{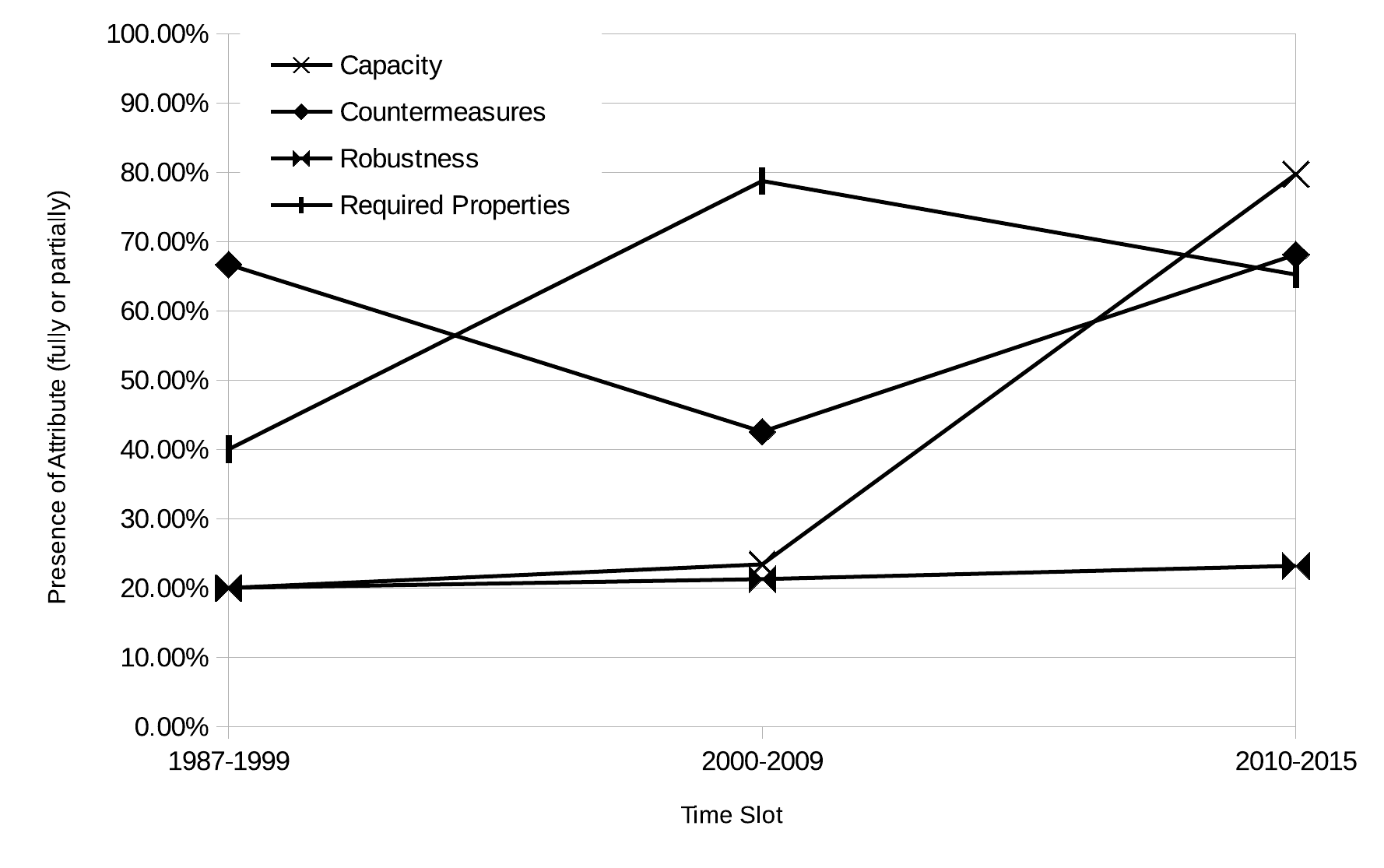}
  \caption{Coverage of selected attributes for hiding methods over time.}
  \label{Fig:attribDevel}
\end{figure}

The attributes covered by publications changed over time. Figure~\ref{Fig:attribDevel} provides an overview of selected attributes over time. We omitted the sender-side and receiver-side processes that were described in most publications but in very varying detail.

Over time, the covert channel capacity was increasingly discussed, especially in publications from the last six years (2010--2015). Channel robustness was constantly discussed for only 20\% of the hiding methods. Both, channel capacity and robustness are part of the `Covert Channel Properties' in our description method. The discussion of countermeasures varied over time and is similar in the range 2010--2015 as it was in 1987--1999 (approx. 68\%). The required properties of the carrier were discussed by fewer publications in recent years (2010--2015) compared to the years 2000--2009.

\begin{figure}[!th]
\centering
  \includegraphics[width=0.99\textwidth]{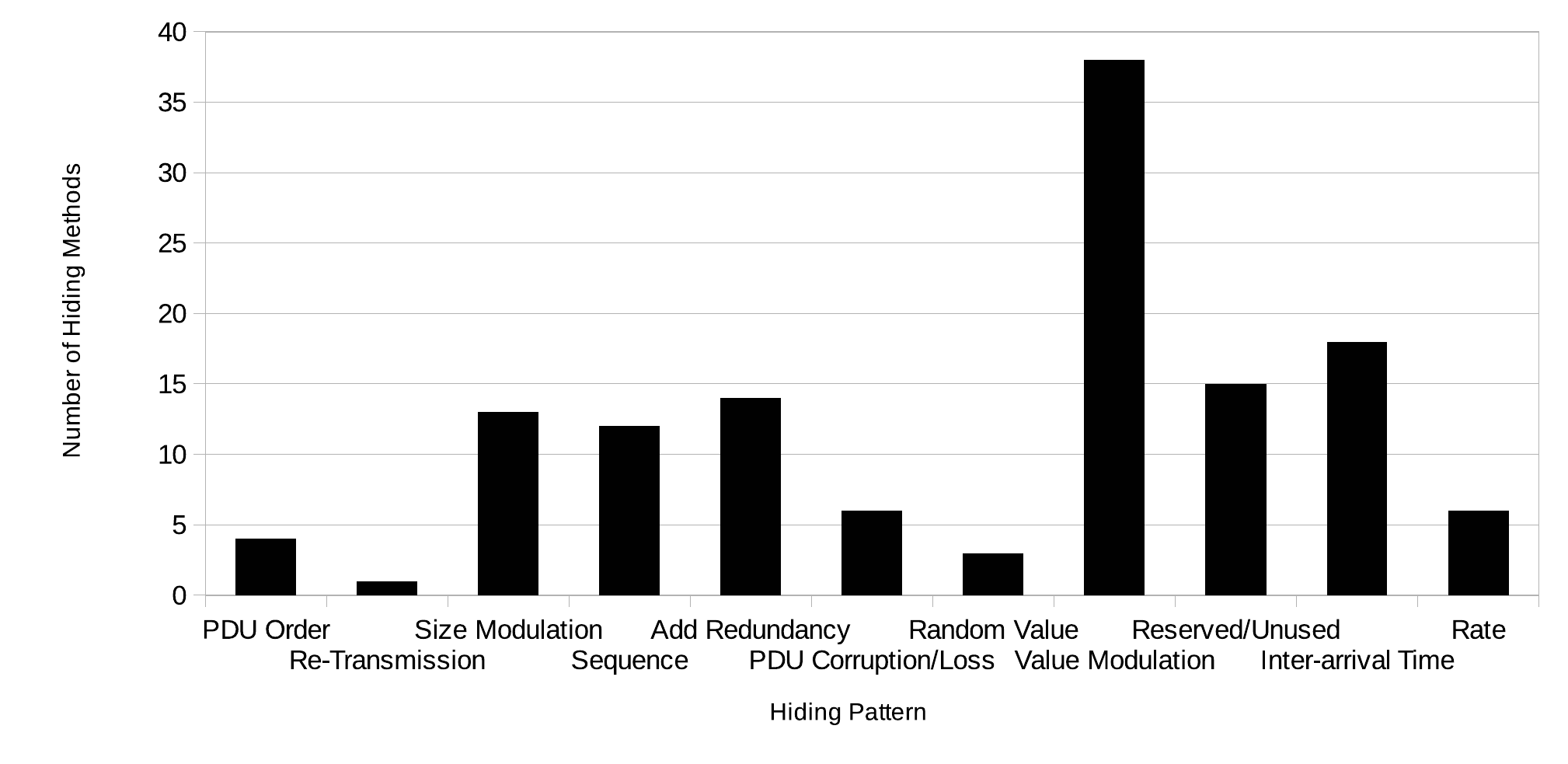}
  \caption{Occurrences of hiding patterns for the analyzed hiding methods.}
  \label{Fig:hidPatterns}
\end{figure}

As hiding patterns were proposed only recently, none of the existing publications covered hiding patterns. We analyzed all hiding methods by verifying whether they were already assigned a pattern in \cite{Wendzel:CSUR} and for several methods which had not been analyzed in \cite{Wendzel:CSUR}, we determined their hiding patterns. We were able to assign 130 of the 131 analyzed hiding methods to their respective patterns. One publication discussed a steganographic key exchange that applies to many hiding methods and thus cannot be assigned to a hiding pattern. Figure~\ref{Fig:hidPatterns} shows the distribution of hiding patterns.

\subsection*{\textbf{Finding 3: Inconsistent descriptions of methods within the same paper}}

Some of the publications describe several hiding methods while applying an inconsistent description for these. We use \cite{CCinPervasiveComputing} as an example for this scenario. The authors present two covert channels for pervasive computing environments. The first signals hidden information by modulating the Link Quality Indication (LQI) of 802.15.4 wireless networks while the second modulates the values of a temperature sensor to signal hidden information. Table~\ref{tab:MultChanDescDiffInOnePaper} indicates which of the selected attributes of the description are present, partially present, or not present. The table reveals that different attributes  are described for each channel. The lack of a unified description makes it harder to compare different channels.

\doublerulesep 0.1pt
\begin{table*}[!th]
\centering
\begin{footnotesize}
\caption{Inconsistent descriptions and combined discussions of attributes within papers: present attributes of the hiding methods described in~\cite{CCinPervasiveComputing} and~\cite{SDPbasedCC}} \label{tab:MultChanDescDiffInOnePaper}
\begin{tabular}{p{2.25cm}p{1.3cm}p{1.2cm}p{1.4cm}p{1.5cm}p{1.2cm}p{1.5cm}p{1.5cm}}
\hline\hline\noalign{\smallskip}
    Hiding Method & Application Scenario & Required Properties & Counter-measures & Sender-Receiver Relation & Direct/ Indirect & Robustness & Capacity\\
\noalign{\smallskip} \hline
    Link quality~\cite{CCinPervasiveComputing}      & \Yes,combined & \Par & \Yes & \No  & \No  & \Yes & \Yes \\
    Sensor data~\cite{CCinPervasiveComputing}	    &               & \Par & \Par & \Yes & \Par & \Yes & \Par \\
\hline
    SDP o-tag~\cite{SDPbasedCC}  & \Yes,combined  & \Par,combined   & \No  & \No   & \No   & \Par,combined  & \Par,combined  \\
    SDP a-tag~\cite{SDPbasedCC}  & 		      &                 & \No  & \No   & \No   &                &                \\
\hline\hline
\end{tabular}
\end{footnotesize}
\end{table*}

\subsection*{\textbf{Finding 4: Several papers combined the evaluation of multiple hiding methods}}

We also found that sometimes multiple hiding methods are treated in a combined way throughout a paper.  Tab.~\ref{tab:MultChanDescDiffInOnePaper} summarizes the combined and the non-combined attributes of~\cite{SDPbasedCC}. The authors describe two hiding methods for SIP useful for the stealthy command-and-control communication in a botnet. The first channel uses the mandatory `<o>' tag to carry hidden information while the second uses the optional `<a>' tag and its parameter to do the same. Both channels are combined for the evaluation process as both channels are necessary to transfer the required secret message. This combined evaluation does not allow the reader to understand the performance of each channel separately and it also makes the comparison against other methods difficult.

\section{Linking the Description Method and the Creativity Framework}\label{Sect:Creativity}

In the creativity framework \cite{StegoCreativity}, the major focus is on the evaluation of creativity, especially originality, of a proposed hiding method. 
We will briefly describe the creativity framework and afterwards explain how the new unified description method can be used to improve it.

\emph{Steps of the Creativity Framework:} The creativity framework consists of five steps which are aligned with the traditional peer review process. In step one, a pattern database is generated by the research community. Due to the availability of a pattern catalog \cite{Wendzel:CSUR,NetStegBook}, step one is already accomplished.

In the second step, the authors create the idea of a new hiding method, e.g.\ to embedd hidden bits into a new network protocol. The authors describe their new hiding method in form of a scientific paper in step three. They justify the novelty of their method using the \emph{creativity metric} \cite{StegoCreativity}, which is based on the originality of the method (hiding method pattern) as well as the context (application scenario and carrier network protocol).

The authors submit their paper to a peer review. The reviewers evaluate the novelty of the work using the creativity metric and decide whether the proposed work is a ``Big-C'' or a ``small-c'' contribution, i.e.\ whether the work consists a high level of creativity. Big-C and small-c are standard terms from creativity research. Only in the Big-C case, the work is accepted to represent a new pattern and its pattern description is optimized in step four, otherwise step four is skipped.

\pagebreak
The work is published in step five. In case of ``Big-C'' the publication has to state that the work represents a new pattern -- this automatically extends the pattern database. In case of ``small-c'', the hiding method is published but the authors cannot claim that they have discovered a new pattern; instead they provide new results for an existing hiding pattern.

\emph{Benefits of Combining Both Approaches:}
The creativity metric does not enforce a detailed description structure. Our new unified description method can replace the creativity metric, since it provides a more fine-grained description and allows for the distinction and comparison of various aspects of a hiding method. Several attributes, such as whether a hiding method can operate in a MitM setup or a distinction between the sender-side and receiver-side processes, were not covered in \cite{StegoCreativity}. 
On the other hand, our unified description method can benefit from the creativity framework. There is no reason to create a new approach for integrating the unified description method into the peer review process. Also, the creativity framework evaluates the novelty of a hiding method by applying research from creativity psychology, which can serve as an add-on to the technical evaluation of the unified description method. 
In summary, the combination of both approaches has many benefits. 

\section{Discussion and Conclusion}\label{Sect:Discuss}

We developed an approach to unify and structure the description of network steganographic methods. To this end, we performed a comprehensive literature analysis in the domain to identify requirements for the description method. Currently, no such description exists, making it difficult to compare the published work on hiding methods. Unified descriptions of hiding methods are desirable as they ease the comparison of research results. They also improve the accessibility of hiding methods and foster the reproduction of experimental results. As a key aspect of our description method is the association of hiding methods with a hiding pattern, the approach automatically enforces a categorization of the research results.

However, unified descriptions for hiding methods are hard to enforce as, in the end, it is up to every author to decide how to describe his or her hiding method. For this reason, we designed our description so that it will be applicable and attractive for many authors. Our description structure does not specify every single detail of all attributes. It leaves several decisions to the authors, such as whether or not to discuss details of certain attributes (e.g.\ covert channel capacity), the form of descriptions (e.g.\ text or figures), and the extent of the descriptions. This flexibility allows to apply the unified description method also in short papers which are, for many conferences, limited to four to six pages. As some attributes are specified as `optional', they can be also left out.

We propose combining the new unified description method with the existing `creativity framework'. The framework's process is kept but the `creativity metric' of the framework is replaced with our unified description method.

\bibliographystyle{unsrt}
\bibliography{bmc_article}

~

\footnotesize
\textbf{Steffen Wendzel} is head of a research team on smart building security at the Fraunhofer Institute for Communication, Information Processing and Ergonomics (FKIE), Bonn, Germany. He studied Computer Science in Kempten, Augsburg and Hagen and received his PhD in 2013 from the University of Hagen. He published several papers on network information hiding and authored five books. He is also a reviewer for several major journals and conferences and gave more than 50 talks. Steffen's research interests are network covert channels, building automation systems, and information security terminology. He is also a developer of several open source software projects. His website is \emph{https://steffen-wendzel.blogspot.de}.

~

\textbf{Wojciech Mazurczyk} holds M.Sc.\ (2004), Ph.D.\ (2009, with honours) and D.Sc.\ (habilitation, 2014) all in Telecommunications from Warsaw
University of Technology (WUT), Poland, where he currently works as Associate Professor. He is the author of more than 100 scientific 
papers, and more than 30 invited talks on information security and 
telecommunications. Head of Bio-inspired Security Research Group (\emph{http://bsrg.tele.pw.edu.pl/}) at the Institute of
Telecommunications, WUT. His research interests include bio-inspired
cybersecurity and networking, information hiding and network security. He is also a TPC member of a number of refereed conferences,
including IEEE INFOCOM, IEEE GLOBECOM, IEEE ICC and ACSAC. He also serves as the reviewer for a number of major refereed international
magazines and journals. From 2013 he is an Associate Technical Editor for the IEEE Communications Magazine, IEEE Comsoc. He is a senior 
member of IEEE.

~

\textbf{Sebastian Zander} received a Dipl.-Ing.\ degree in applied computer science 
from Technical University Berlin, Germany in April 1999 and a PhD in 
telecommunications engineering from Swinburne University of Technology, 
Australia in 2010. Currently he is a Lecturer at Murdoch University, Australia. 
Prior to joining Murdoch University, Sebastian worked as a Research Fellow 
and Lecturer at the Centre for Advanced Internet Architectures,
Swinburne University of Technology and as a researcher and project 
manager at Fraunhofer FOKUS, Germany. Sebastian's research interests cover 
several aspects in the area of IP networking and network security, in 
particular covert channels, network traffic classification, network measurement 
and the IPv4 to IPv6 migration. Sebastian is the co-author of over 45 
published journal and conference papers as well as two IETF RFCs. He is also 
the author of several open source software packages. He serves as TPC member 
on several conferences and workshops, including IEEE LCN, and as a reviewer 
for several top-tier ACM and IEEE journals. He is a member of the IEEE and the
ACS. Based on Google Scholar, Sebastian has more than 2,500 citations and an 
h-index of 20 (\emph{https://scholar.google.com.au/citations?hl=en\&user=8csOslsAAAAJ}).

\end{document}